# Title:

**The effect of speech and noise levels on the quality perceived by cochlear implant and normal hearing listeners**

**Authors:**

**Sara Akbarzadeh[1][†]**

[1]Erik Jonsson School of Engineering and Computer Science, University of Texas at Dallas 800 West Campbell Road, Richardson, TX, 75080, USA.

EMAIL: Sara.Akbarzadeh@utdallas.edu

PHONE : (469) 231-5034

[†]Corresponding author

**Sungmin Lee[2]**

[2]Erik Jonsson School of Engineering and Computer Science, University of Texas at Dallas 800 West Campbell Road, Richardson, TX, 75080, USA.

EMAIL: Sung.Lee@UTDallas.edu

**Fei Chen[3]**

[3]Department of Electrical and Electronic Engineering, Southern University of Science and Technology, Shenzhen, China.

EMAIL: fchen@sustech.edu.cn

**Chin Tuan-Tan[4]**

[4]Erik Jonsson School of Engineering and Computer Science, University of Texas at Dallas 800 West Campbell Road, Richardson, TX, 75080, USA.

EMAIL: Chin-Tuan.Tan@utdallas.edu

**Acknowledgments of support:**

**This research was supported by STARs (Science and Technology Acquisition and Retention) program. The authors extend thanks to the subjects who participated in this study.**

# ABSTRACT

Electrical hearing by cochlear implants (CIs) may be fundamentally different from acoustic hearing by normal-hearing (NH) listeners, presumably showing unequal speech quality perception in various noise environments. Noise reduction (NR) algorithms used in CI reduce the noise in favor of signal-to-noise ratio (SNR), regardless of plausible accompanying distortions that may degrade the speech quality perception. To gain better understanding of CI speech quality perception, the present work aimed investigating speech quality perception in a diverse noise conditions, including factors of speech/noise levels, type of noise, and distortions caused by NR models. Fifteen NH and seven CI subjects participated in this study. Speech sentences were set to two different levels (65 and 75 dB SPL). Two types of noise (Cafeteria and Babble) at three levels (55, 65, and 75 dB SPL) were used. Sentences were processed using two NR algorithms to investigate the perceptual sensitivity of CI and NH listeners to the distortion. All sentences processed with the combinations of these sets were presented to CI and NH listeners, and they were asked to rate the sound quality of speech as they perceived. The effect of each factor on the perceived speech quality was investigated based on the group averaged quality rated by CI and NH listeners.

Consistent with previous studies, CI listeners were not as sensitive as NH to the distortion made by NR algorithms. Statistical analysis showed that the speech level has significant effect on quality perception. At the same SNR, the quality of 65 dB speech was rated higher than that of 75 dB for CI users, but vice versa for NH listeners. Therefore, the present study showed that the perceived speech quality patterns were different between CI and NH listeners in terms of their sensitivity to distortion and speech level in complex listening environment.

# 1. INTRODUCTION

Cochlear implants (CIs) are hearing assistive devices from which a growing number of hearing impaired (HI) individuals receive invaluable benefits. These electronic devices deliver acoustic information by using systematically schemed ways (e.g., signal processing strategy, place of stimulation, and stimulation rate) to stimulate auditory nerve fibers along the cochlear. Although its unique hearing restoration mechanisms lead to substantial improvements in speech intelligibility, perceptual details may be different from normal hearing (NH) listeners who use acoustic hearing (Henry et al., 2005; Chatterjee et al., 2010; Goldsworthy et al., 2013; Feng and Oxenham, 2018). In present study, we attempted to examine the effect of diverse speech in noise factors (speech/noise levels, type of noise, distortions caused by noise reduction (NR) models) on the speech quality perception by CI and NH listeners.

Due to the degraded spectro-temporal resolution imposed by technological and physiological limitations in electrical hearing for CI listeners, their speech perception ability is poorer than NH listeners at the same speech level as well as at the same signal-to-noise ratio (SNR). Especially, increased susceptibility to noise is one of the major challenges of CIs. Among many types of background noises, competing speech which fluctuates dynamically and continuously requires a top-down approach to integrate and restore missing speech portion by maskers. This phonemic restoration is known to be hard for CI users in comparison with NH listeners (Nelson and Jin, 2004; Chatterjee et al., 2010; Bhargava et al., 2016). The noise susceptibility is likely to be attributed to poor spectral resolution of CIs caused by the lack of neural survival, limited number of channels, and their interactions (Nelson et al., 2003; Fu and Nogaki, 2005).

In addition to the spectral aspects of CI mediated listening, loudness is not perceived in the same way as that in acoustic hearing. The electrical dynamic ranges of speech processor in CIs are narrower (30-80 dB depending on signal processing schemes and manufacturers) than the acoustic dynamic range of typical sound with normal hearing listener (~120 dB). The wide dynamic range of acoustic inputs are greatly compressed (Skinner et al., 1997; Zeng and Galvin, 1999; Zeng et al., 2002) in CIs to focally deliver the loudness range of conversational speech signals which vary over a 30 dB range. This compressed dynamic range is likely to have negatively influence on speech intelligibility considering the findings of positive relationship between wider input dynamic range and speech perception scores (Holden et al., 2007; Dawson et al., 2007). It has also been reported that CI users require significantly higher level than did NH listeners to reach "comfortable" and "soft" perceptual loudness levels (Luo et al., 2014). Given these unequal input dynamic ranges and distinct auditory mechanisms between CI and NH listeners, there is a possibility that these two groups' perceived quality of speech may differ at the same level or in the same SNR conditions. Our pilot study (Akbarzadeh et al., 2018) examined this question and found an implication that CI users likely prefer softer conversational speech levels in quality rating tasks. On the contrary, NH group rated higher for the speech presented at higher conversational speech levels.

In general, NH individuals prefer higher levels when they are given some conversational levels of sound. The effects of different frequency responses and sound levels on the perceived quality by NH listeners were investigated by Gabrielsson et al (1990). They employed three types of sound set to approximately natural level: pink noise at 68 dB SPL, female voice at 56 dB SPL, and jazz music at 80 dB SPL. For comparison, each sound was also presented at 10 dB lower SPL. The results showed that the higher natural sound level provides better clarity, more fullness, spaciousness, and nearness, but less gentleness than the 10 dB lower SPL. In a part of speech intelligibility study by Hagerman (1982), he showed increase in speech intelligibility until speech levels reach 53 dB HL, but the intelligibility scores started to drop from that point when SNR was held constant. To our knowledge, speech quality rating in response to different levels of speech in noise has not been conducted in CI group.

Presence of noise in speech, certainly increases the errors in identifying and conceiving the meaning of speech. Most of the NR techniques in the front-end processing for hearing devices enhance the speech signal in favor of SNR. Some studies showed evidence that the adverse effects of NR algorithms on sound quality perception are worse than the negative effect of background noise (Kates, 1993). Baer et al (1993) conducted a series of experiments to evaluate the effect of digital processing of noisy speech on intelligibility and quality for HI subjects. They found that large amount of speech enhancement decreases the intelligibility of speech in noise. Subjects' performance for moderate degree of enhancement was generally the same as that for no enhancement. It seems a doubtless fact that most NR algorithms cause nonlinear distortions that degrade the quality of speech signals.

In this study, we aimed to investigate perceptual difference between acoustic and electric hearing by comparing the speech quality in noise ratings between NH and CI groups. Two types of noise were used, i.e., babble and cafeteria noise. Speech and noise levels were varied with an assumption that auditory perception of noisy speech not only relied on SNR level, but also on the level of speech above audition. Two NR schemes were applied to generate nonlinearly distorted speech allowing us to identify sensitivity to degraded sounds for two groups.

## 2. METHOD

### 2.1. Participants

Demographic details of CI subjects are shown in Table 1. Fifteen NH subjects (8 males and 7 females; mean age of 21 years, and a standard deviation of 3.3 years) and 7 CI users (2 males and 5 females, mean age of 53 years, and a standard deviation of 20.77 years) participated in this study. Hearing screening was conducted for each NH subject with 20 dB HL tones across octave frequencies to verify their NH. For CI subjects, aided- and unaided- audiometry was carried out for each individual. Their hearing thresholds were identified to be in the range of profound-to-severe hearing loss or deaf. Their aided auditory thresholds were better than 35 dB HL for all octave frequencies from 500 Hz to 4 kHz. All CI users had more than one year of CI experience. All subjects were native speakers of American English. They completed informed consent prior to the experiment and they were compensated for their participation. All the procedures were approved by University of Texas at Dallas Institutional Review Board.

**Table 1- Demographic details of CI participants.**

| N | Gender | Age (years) | Etiology | CI Manufacturer and model | Uni/ Bi-lateral CI |
|---|---|---|---|---|---|
| 1 | Male | 65 | Noise exposure | Cochlear Nucleus 6 | Bilateral |
| 2 | Female | 61 | Hereditary | Cochlear Nucleus 6 | Unilateral |
| 3 | Male | 68 | Unknown | Cochlear Nucleus 6 | Bilateral |
| 4 | Female | 24 | Meningitis | Cochlear Nucleus 6 | Unilateral |
| 5 | Female | 61 | Hereditary | Cochlear Nucleus 6 | Unilateral |
| 6 | Female | 22 | Ototoxicity | Cochlear Nucleus 6 | Unilateral |
| 7 | Female | 70 | Ototoxicity | Cochlear Nucleus 6 | Unilateral |

## 2.2. Stimulus

Four original speech sentences were produced by randomly concatenating two short sentences spoken by male and female speaker extracted from AzBio database (Spahr et al., 2012). These sentences were set at two sound pressure levels (SPL): 65 dB and 75 dB SPL to make clean sentences. To create noisy sentences, two types of noise were chosen: Cafeteria and Babble noise. The noises were set to three different levels: 55, 65, and 75 dB SPL. By adding up the noises to clean sentences in all possible conditions, the noisy sentences were created. Therefore, altogether we generated 8 clean sentences (4 original sentences × 2 speech levels) and 48 noisy speech sentences (4 original sentences × 2 speech levels × 3 noise levels × 2 noise types). The speech stimuli were further processed using two classical NR algorithms: NR1-Wiener filtering method (Lim and Oppenheim, 1978), NR2-Binary Masking method (Hu and Loizou, 2004), and No-NR-Sentences that were not processed by either of these two algorithms. Totally, our speech dataset consists of 24 sentences (4 sentences × 2 speech levels × 3 NRs) in quiet condition, and 144 sentences (4 sentences × 2 speech levels × 3 noise levels × 2 noise types × 3 NR) in noisy condition.

## 2.3. Procedure

Stimuli were produced and controlled in MATLAB. The stimuli were presented via a 24-bits Soundcard (RME Fire face UC) through a pair of headphones (Sennheiser HD 600) for NH subjects, and through a loud speaker (AURATONE 5C) for CI subjects. For subjects with CI, they used programming set on their CI that they use in daily life. Subjects were seated in a sound treated booth with a monitor at 1 meter in front of them. Subjects' task was to rate the quality of the presented sentences. At the onset of each stimulus, a rating bar numbered from 1 to 10, 1 representing the most unnatural and 10 representing the most natural, was appeared on the monitor. Subjects were conducted to provide their rating by clicking on the number tab. After each stimulus was presented, the program waits infinitely for the listener to rate the sound quality. Repetition of the last stimulus was given as an option to the listener. The subjects were especially encouraged to focus on the target speech, and rate them, rather than rating overall sound inputs including background noise. The experiment was conducted twice for each listener to identify consistency on their responses. The two sessions lasted approximately an hour to complete.

## 3. RESULTS

### 3.1. Consistency of quality rating

Each subject rated the randomized stimulus set twice in two separate trials. To quantify how consistent the subjects were in their rating, the Pearson correlation between these two trials for each subject was calculated. The inter-trial correlations for all subjects were high ($r$=0.87-0.99), indicating the high consistency in evaluating the quality of speech.

### 3.2. Quality rating in quiet

The grand average of quality ratings by NH and CI subjects in quiet condition (0 dB SPL noise) is depicted in Figure 1. The rating results were extremely high (>0.9) across all the conditions for both NH and CI group. Due to the considerable ceiling effect, further analysis was not carried out.

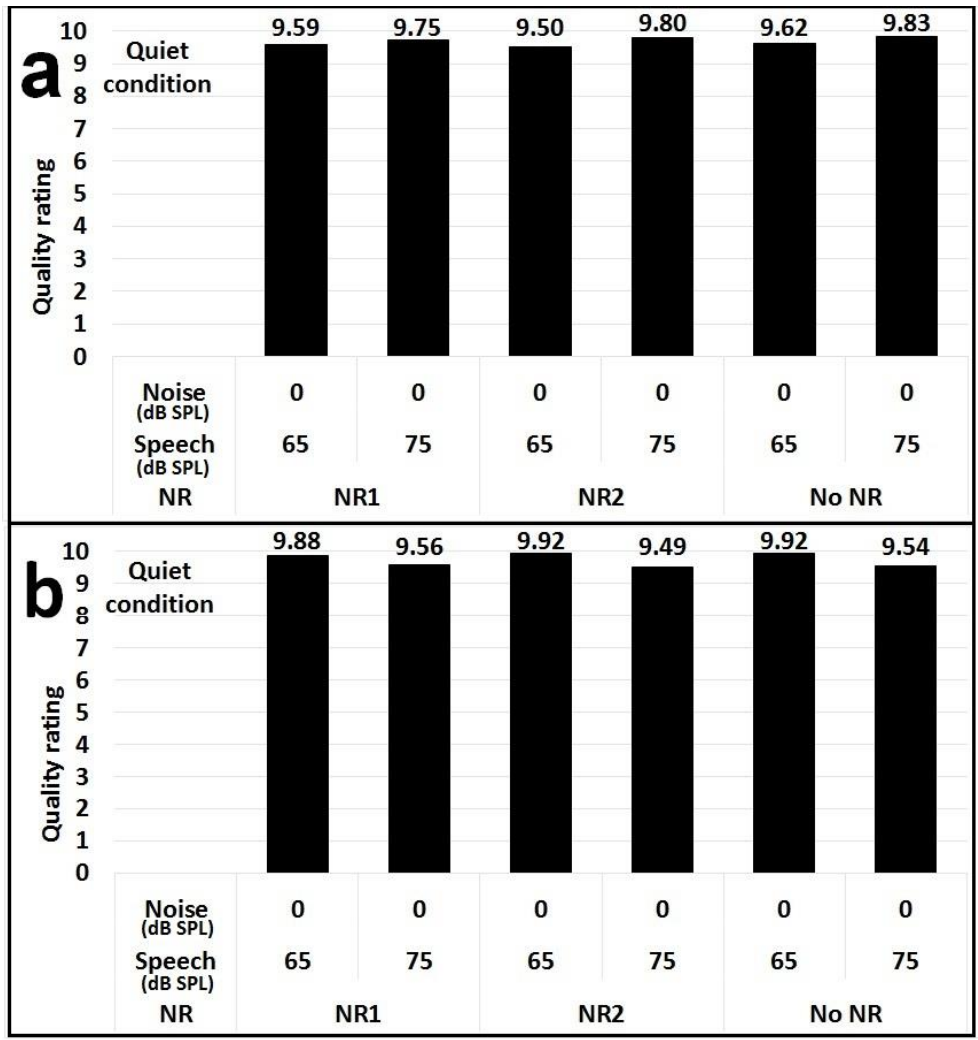

**Figure 1- Grand average ratings in quiet condition by a) NH, and b) CI subjects**

### 3.3. Quality rating in noise

Figure 2a and b represent the grand average quality rating of the speech in Babble and Cafeteria noise for NH and CI listeners respectively. We observed a systematic trend in rating the quality perceived by both NH and CI listeners. For each NR processing algorithm, at the same speech level, the quality rating decreases as the noise level increases. In similar way, at the same noise level, the quality rating increases as the speech level increases.

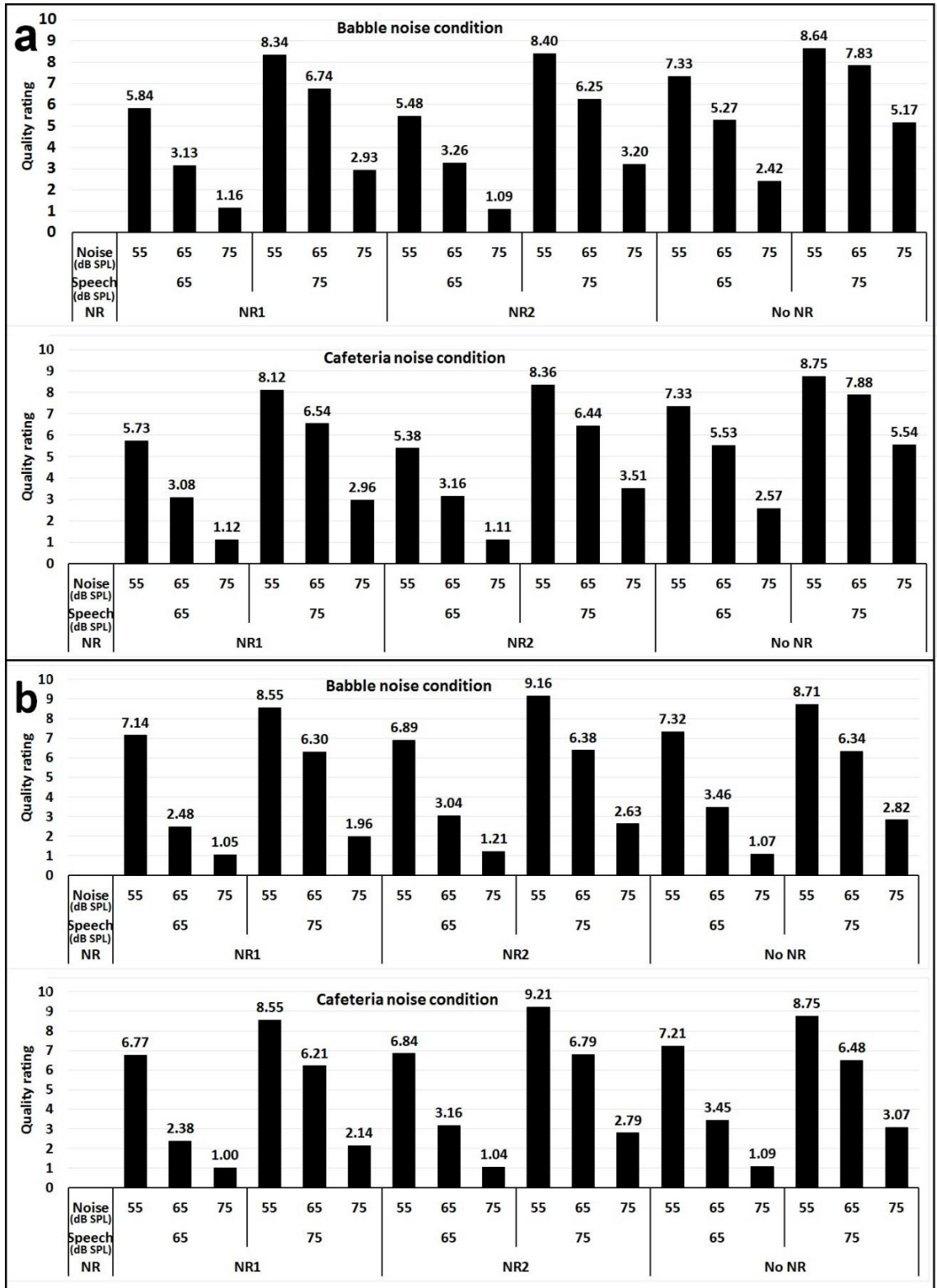

**Figure 2- Grand average ratings in Babble and Cafeteria noise by a) NH, and b) CI subjects.**

A mixed analysis of variance (ANOVA) was conducted on quality rating to figure out whether quality rates systematically vary based on the independent variables. The ANOVA had a between-subject factor of hearing status (NH or CI) and within-subject factors of noise type (Babble or Cafeteria), NR algorithm (NR1, NR2, and No-NR), speech level (65 and 75 dB SPL) and noise level (55, 65 and 75 dB SPL). The results indicates that there was a significant main effect of hearing status, NR algorithm, speech level and noise level, but no main effect of noise type (Table 2). A pairwise comparison with Bonferroni adjustment showed that there is a significant difference between all pairs of noise levels at 55, 65 and 75 dB SPL, representing lower quality rates associated with higher noise levels scores ($p<0.05$). Further post hoc test using Bonferroni corrections revealed that the quality rating for No-NR condition is significantly better than the quality rating for NR1 or NR2 ($p<0.05$), but there was no significant difference between NR1 and NR2. There was significant interaction of the hearing status with NR processing algorithm, indicating that NH listeners rated higher than CI users for No-NR and NR1 conditions, while CI listeners rated higher for NR2. Another significant interaction between hearing status and noise level represented that NH rated higher when noise was either 65 or 75 dB SPL, but CI rated higher than NH when noise was at 55 dB SPL. Significant interactions were also found for speech level * noise level and NR algorithm * speech level * noise level. Overall, it appears that the acoustically manipulated variables and group variables affect quality ratings with some extents of interplay between each other.

**Table 2- Results of mixed ANOVA conducted on perceived quality rating with factors of Hearing status(NH and CI), Noise type (Babble and Cafeteria), Processing algorithm (NR1, NR2 and No-NR), Speech level (65 dB and 75 dB), and noise level (55 dB, 65 dB and 75 dB).**

| Factor | df | F | *p* |
|---|---|---|---|
| **HS (Hearing Status)** | 1 | 8.692 | 0.003* |
| **NT (Noise Type)** | 1 | 0.059 | 0.808 |
| NR (Noise Reduction) | 2 | 28.386 | < 0.001* |
| SL (Speech Level) | 1 | 437.029 | < 0.001* |
| NL (Noise Level) | 2 | 730.778 | < 0.001* |
| HS*NT | 1 | 0.003 | 0.958 |
| HS*NR | 2 | 15.451 | < 0.001* |
| HS*SL | 1 | 1.712 | 0.191 |
| HS*NL | 2 | 17.742 | < 0.001* |
| NT*NR | 2 | 0.266 | 0.766 |
| NT*SL | 1 | 0.356 | 0.551 |
| NT*NL | 2 | 0.191 | 0.826 |
| NR*SL | 2 | 1.354 | 0.259 |
| NR*NL | 4 | 0.920 | 0.451 |
| SL*NL | 2 | 13.887 | < 0.001* |
| HS*NT*NR | 2 | 0.039 | 0.962 |
| HS*NT*SL | 1 | 0.106 | 0.745 |
| HS*NT*NL | 2 | 0.027 | 0.973 |
| HS*NR*SL | 2 | 0.157 | 0.854 |
| HS*NR*NL | 4 | 0.355 | 0.841 |
| HS*SL*NL | 2 | 2.414 | 0.090 |
| NT*NR*SL | 2 | 0.042 | 0.959 |
| NT*NR*NL | 4 | 0.028 | 0.998 |
| NT*SL*NL | 2 | 0.047 | 0.954 |
| NR*SL*NL | 4 | 2.907 | 0.021* |
| HS*NT*NR*SL | 2 | 0.023 | 0.977 |
| HS*NT*NR*NL | 4 | 0.031 | 0.998 |
| HS*NT*SL*NL | 2 | 0.007 | 0.993 |
| HS*NR*SL*NL | 4 | 0.225 | 0.924 |
| NT*NR*SL*NL | 4 | 0.028 | 0.999 |
| HS*NT*NR*SL*NL | 4 | 0.017 | 0.999 |

### 3.4. Quality rating at 10 dB SNR

To tease out the effect of SNR, and investigate the effect of speech and noise levels alone, the average ratings for noisy speech at 10 dB SNR were extracted from Figure 2, and depicted in Figure 3. Opposite trends in quality ratings between NH and CI subjects are illustrated in Figure 3. All NH subjects rated higher for the 10 dB SNR conditions with louder speech (75 dB SPL) in noise (65 dB SPL) (Fig. 3a), whereas CI subjects rated higher for the conditions with softer speech (65 dB SPL) in noise (55 dB SPL) (Fig. 3b). This indicates two groups may perceive sound quality differently depending on the given sound level.

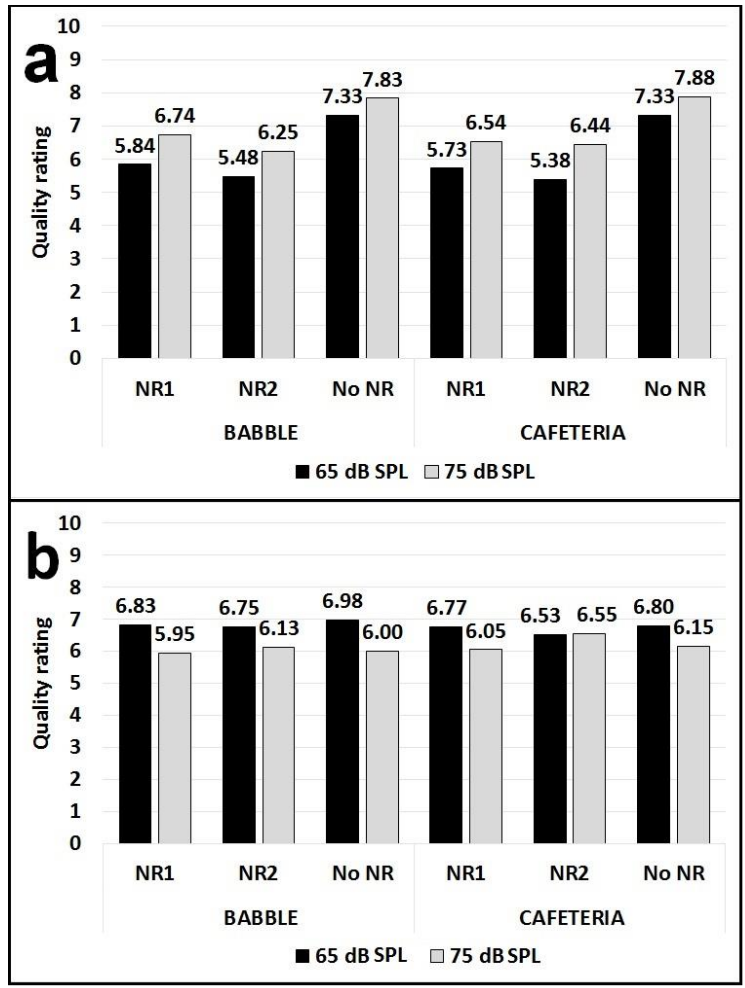

**Figure 3- Grand average quality rating in 10 dB SNR with the speech level at 65 dB and 75 dB for a) NH and b) CI subjects.**

A mixed ANOVA was conducted for average quality rating at 10 dB SNR with between subject factors of hearing status and within factors of noise type, processing algorithm and speech level. The results show the significant main effect of NR algorithm [$F(2,240)=112.893, p=0.002$]., but no significant effect of the other factors: group [$F(1,240)=0.474, p=0.492$], noise type [$F(1, 240)=0.008, p=0.929$], and speech level [$F(1, 240)=0.332, p=0.741$]. Significant interaction effects were only found between the hearing status * NR algorithm [$F(2,240)=4.232$, $p=0.016$] and hearing status * speech level [$F(1,240)=8.949$, $p=0.003$].

In order to interpret the interaction effects, we run ANOVA with factors of noise type, NR processing, and speech level separately for NH and CI subjects (Table 3 and Table 4). The results for NH subjects showed the significant main effects of speech level and NR algorithm. That is, the perceived quality rating by NH subjects at 75 dB SPL speech is significantly higher than that at 65 dB SPL speech. A pairwise comparison for NR processing showed that No-NR is significantly higher than NR1 and NR2, but there is no difference between NR1 and NR2. Table 4 represents the significant main effect of speech level on perceived quality for CI subjects. As we see in Figure 3b, the statistic revealed that perceived quality rating by CI subjects at 65 dB SPL speech is significantly higher than that at 75 dB SPL. Unlike NH subjects, NR processing has no significant effect on quality rating for CI subjects.

**Table 3- Results of mixed ANOVA conducted on perceived quality rating by NH subjects at 10 dB SNR with factors of Noise type (Babble and Cafeteria), Processing algorithm (NR1, NR2 and No-NR), and Speech level (65 dB and 75 dB). (For abbreviations refer to Table 2)**

| Factor | df | F | p |
|---|---|---|---|
| NT | 1 | 0.223 | 0.644 |
| NR | 1.233 | 15.690 | 0.001* |
| SL | 1 | 6.559 | 0.023* |
| NT*NR | 2 | 0.931 | 0.406 |
| NT*SL | 1 | 0.202 | 0.660 |
| NR*SL | 1.577 | 1.062 | 0.348 |
| NT*NR*SL | 1.255 | 0.167 | 0.744 |

Table 4- Results of mixed ANOVA conducted on perceived quality rating by CI subjects at 10 dB SNR with factors of Noise type (Babble and Cafeteria), Processing algorithm (NR1, NR2 and No-NR), and Speech level (65 dB and 75 dB). (For abbreviations refer to Table 2)

| Factor | df | F | p |
| --- | --- | --- | --- |
| NT | 1 | 0.030 | 0.868 |
| NR | 1.008 | 0.102 | 0.763 |
| SL | 1 | 25.693 | 0.002* |
| NT*NR | 2 | 1.033 | 0.386 |
| NT*SL | 1 | 5.034 | 0.066 |
| NR*SL | 2 | 1.607 | 0.241 |
| NT*NR*SL | 2 | 0.115 | 0.892 |

## 4. DISCUSSION

The purpose of this study was to examine the speech quality perception for acoustic hearing and electrical hearing by comparing NH and CI listeners. To this end, a diverse set of speech in noise conditions were designed by applying the combinations of speech and noise levels, NRs, and type of noise. Generally, the average quality was rated higher at higher SNR for both NH and CI listeners. Speech quality ratings for CI users differed from those for NH listeners in some aspects of perceiving degraded speech in noise environments.

Previous studies showed that inter-trial reliability for HI subjects are lower than that for NH subjects (Tan and Moore, 2008; Narendran and Humes, 2003). However, our results showed high inter-trial correlations ($r = 0.87$-$0.99$) for both groups, indicating that CI subjects were as consistent as NH subjects in rating the perceived quality of speech. Presumably, the stimulus sets designed in this study would be apparently distinct enough for CI users to judge the quality in a ten rating scales.

In quiet conditions, we observed the ceiling effect that exhibits considerably high rates across all NR processing and speech level conditions. It would make sense that the quality rating was extremely high for quiet conditions because NR algorithms were not operated in quiet, and distortions did not occur to speech. However, it may also be possible that our subjects rated high for those quiet conditions, due to the absence of background noise which may facilitate perceiving speech signal clearly. Despite our encouragement for them to pay attention to the clarity and intelligibility of target speech, not noise, our subjects may have felt harder to segregate speech from noise and evaluate whole stimuli in noise conditions.

The interesting part of our results is the comparison between the quality rating of the sentences at 10 dB SNR by NH and CI subjects. When SNR was held constant at 10 dB, NH subjects rated higher for 75 dB SPL speech over for 65 dB SPL speech. On the other hand, the CI subjects' rating was higher for 65 dB SPL than 75 dB SPL. Assuming that the quality of speech is maximized with speech at the most comfortable levels, we expected that CI users would prefer speech with higher conversational level (75 dB SPL) over lower level (65 dB SPL), as they exhibited softer sound perception (Luo et al., 2014) with narrower input dynamic range (Hong et al., 2003), compared to NH listeners. In contrast to our expectation, our CI subjects rated higher for lower level speech. We speculate that higher noise level accompanied with the higher speech level may sound uncomfortably noisy for our CI listeners, so they gave higher rate to the lower speech at the cost of the speech audibility. This assumption is associated with the findings that CI users less tolerate noise, showing speech recognition that is more susceptible to background noise than that of NH (Nelson et al., 2003; Mao and Xu, 2017). Our finding is in line with the findings from other CI literatures that shows negative effect of increasing conversational speech levels in noise on speech intelligibility for CI users (Khing et al., 2013).

In our results, the perceived quality of sentences in No-NR condition was significantly higher than that in NR1 and NR2 condition when rated by NH subjects. In the quality rating by CI listeners, however, there was no significant difference between three NR processing conditions. This implies that CI listeners are not as sensitive as NH listeners to the distortion produced by NR algorithms. This is in line with findings from other studies showing higher sensitive perception to degraded speech for NH listeners over HI listeners (Lawson and Chial, 1982; Koning et al., 2015). Poorer spectral- and temporal-resolution of CI listeners may not be capable of recognizing changes in speech quality caused by NRs. Consequentially, the perceptual effect of acoustic distortions elicited by NR techniques differs between NH and HI listeners, so user-dependent NR strategies should be applied in hearing devices. For example, people with sever hearing loss who typically are vulnerable to noise, but not sensitive to distortions, would be able to take advantage of more aggressive NR techniques in their hearing aid or cochlear implant.

## 5. CONCLUSIONS

In summary, this study provided results on the perceived sound quality rating by NH and CI subjects in different speech and noise levels, noise types and NR processing. The results suggested that in addition to SNR, speech level is another important factor that affects the quality rating in NH and CI subjects under background noise. At the fixed SNR, NH subjects preferred in quality for speech at higher conversational level, whereas CI subjects rated higher for softer speech. The difference in quality rating pattern between two groups may be associated with more susceptibility to noise for CI subjects caused by characteristics of electrical hearing. This study gives clear evidence that CI listeners prefer the lower level of speech in lower noise level to the higher level of speech in higher noise level. This implies that they rather choose lower noise at the expense of poorer speech audibility. Knowing the pattern of perceived quality of speech by CI listeners in comparison with NH listeners will help us to improve the signal processing strategy in CIs by making their hearing pattern similar to NH hearing pattern.

*Acoust Soc Am*. 111:377–386.